

\magnification=1200
                       \bigskip

          \centerline {\bf Batalin--Vilkovisky Formalism}
          \centerline {\bf and}
          \centerline {\bf Integration Theory on  Manifolds}

                       \bigskip
                       \bigskip
       \centerline   {{\bf O.M. Khudaverdian}\footnote{$^a$}
    {Supported in part by Suisse Science foundation and by Grant No. M3Z300
of International Science Foundation}}
 \centerline {\it Department of Theoretical Physics, Yerevan State
                                           University}
   \centerline{A. Manoukian St., 375049  Yerevan, Armenia
          \footnote{$^b$}{Permanent address} }
  \centerline {e-mail: "khudian@vx1.yerphi.am"}
            \centerline {and}
 \centerline { \it Department of Theoretical Physics, Geneva University}
    \centerline {1211 Geneva 4, Switzerland}
        \centerline{e-mail: "khudian@sc2a.unige.ch"}
                      \bigskip

        \centerline    {\bf A. Nersessian}

  \centerline   {\it Bogoliubov Laboratory of  Theoretical Physics.
    Joint Institute for Nuclear Research}
    \centerline {\it Dubna, Moscow Region  141980 Russia}
 \centerline  {e-mail: "nerses@thsun1.jinr.dubna.su"}

                          $$
                          $$

     \centerline {UGVA--DPT 1995/07--896}

                       $$ $$
 {\it The correspondence between the BV-formalism and integration theory on
supermanifolds is established.  An explicit formula for the  density on
a Lagrangian surface in a superspace provided with
 an odd symplectic structure and a volume form is proposed.}

  \vfill\eject

  \centerline {\bf 1 Introduction}
                     $$ $$
     In their outstanding works [1]
  Batalin and Vilkovisky
 proposed the most general method for quantizing arbitrary gauge field
 theories.

      During the years it becomes clear that this scheme is  very powerful
 for resolving ghost problems and moreover it contains a rich geometrical
structure . In  the paper [2] Witten proposed a program for the
 construction of String Field Theory in the framework of the Batalin-Vilkovisky
 formalism (BV-formalism) and noted the necessity
 of its geometrical investigation.
 The BV-formalism indeed uses the geometry of
 the superspace provided with odd symplectic structure and the volume
form . The properties of this geometry and its connection to the BV
 formalism was investigated for example, in [3,4,5,6].
 Particularly in [5] A. S.  Schwarz gives the
   detailed geometrical
analysis of the BV-formalism in  terms of this geometry.

   However, some
 specific aspects of the BV-formalism  are not completely clarified
   , such as:

 --  the
   geometrical meaning of the
 initial conditions of the master-action,

  -- the choise of
   the gauge fermion
    and   the geometrical reasons for
   the extending  the initial space of  fields
    with ghosts and antighost
  fields.

  In this work we try to analyze some of these questions.
    For this purpose we study the analogy between the BV--scheme and
  the corresponding constructions in differential geometry.

 From the geometrical point of view to the gauge
 symmetries correspond the vector fields
   on the space of the classical fields which preserve the action.
   The partition function,
   when gauge conditions are fixed, is the integral
  of a nonlocal density constructed by means of these
   vector fields over the
   surface which is defined by gauge conditions.
  This surface is embedded  in the space of the  classical fields.

   The gauge independence  means that this density have to be closed.
 To make this density local in the BV formalism one have to rise
the density and the gauge fixing surface on the extended space:
 to the gauge fixing surface corresponds  the Lagrangian
   manifold embedded in the phase
 space of the "fields" and "antifields" ("fields"= classical fields, ghosts),
    to the closed  density corresponds the volume form on this manifold
 (the exponent of the BV master--action) which obeys
  to BV master equation [1,5,6].

  In the  2-nd Section we briefly recall the basic formulae of the
  BV formalism
 and  following  [5] give the
  covariant explicit formula for the volume element on the
 Lagrangian manifold when it is given by
 arbitrary functions of the fields  and antifields.
 This formula is related to the  multilevel field-antifield formalism
 with the most general Lagrangian hypergauges [11].

  In the 3-th Section we briefly recall the basic constructions
of the geometry of the superspace provided with an odd
 symplectic structure and volume form [3,5,6].---
 It is this geometry on which the BV formalism is based, and
 which development on the other hand was highly inspired by this formalism.
 In particular we shortly describe the properties of the
 $\Delta$-- operator arising in this geometry and the connection
between the BV--formalism and the
 $\Delta$--operator nilpotency condition.

  In the 4-th Section we consider the densities [7,8,9,10]
  (the general covariant objects
 which can be integrated over supersurfaces in the superspace). Following
 [8] and [10]  we consider a special class of densities ---
  pseudodifferential forms
  on which the exterior
 derivative can be defined correctly.  Using Baranov-Schwarz (BS)
 transformations [8] we rise these forms to
integration objects on the enlarged space and formulate the condition
 of closure of these forms in  terms of the $\Delta$--operator.

 In the 5--th Section, using BS transformations
 we study the relations
  between gauge
symmetries in field theory and the closed pseudodifferential forms
 corresponding to the integrand for the partition function of the theory.
We study the relations between the closure conditions and
 the BV--master--equation.

                $$ $$  $$ $$

    \centerline     {\bf 2. BV Formalism}
                       \medskip
       In this section we recall the basic constructions
      of BV formalism [1]: the integral for the partition function
       and we rewrite this integral in the case where the lagrangian
    manifold is given in  covariant way.

     Let $S(\phi)$ be the action of theory with gauge symmetries
 $\{R^A_b (\phi)\}$:
                                $$
             R^A_b (\phi){\delta S(\phi) \over \delta \phi^A}=0\quad .
                                                    \eqno(2.1)
                      $$
 We use de Witt condensed notations (index $A$ runs over the
  all the indices and the spatial coordinates of
   the fields $\phi$).
    Let ${\cal E}$ be the space of the fields $\Phi^A$ and antifields
   $\Phi_{* A}$ where $\Phi^A=(\phi^A, c^b,\nu_b,...)$ is the space of fields
    $\phi^A$ enlarged with the ghosts,lagrangian multipliers for the
constraints e.t.c. and $\Phi^{* A}$ has the parity opposite to
 $\Phi^A$
                           $$
           p(\Phi^{* A})=p(\Phi^A)+1
                                                \eqno(2.2)
                          $$
    In the space ${\cal E}$ one can define the symplectic structure
     by the odd  Poisson bracket:
                               $$
     \{F,G\}={\delta F\over \delta \Phi^A}{\delta G\over \delta \Phi_{* A}}+
             {\delta F\over \delta \Phi_{* A}}{\delta G\over \delta \Phi^A}
                \qquad ({\rm if}\,\, F\,{\rm is}\quad{\rm even})
                                                       \eqno (2.3)
                               $$
   and the $\Delta_0$ operator:
                              $$
              \Delta_0 F=
      {\delta^2 F\over \delta \Phi_{* A}\delta \Phi^A}
                                                     \eqno (2.4)
                              $$
    The master action ${\cal S}$ then can be uniquely defined by the equation
                           $$
                         \Delta_0 e^{\cal S}=
                              0
                           \Leftrightarrow
           \Delta_0 {\cal S}(\Phi^A,\Phi_{* A})+
                      {1\over 2}
      \{{\cal S}(\Phi^A,\Phi_{* A}),{\cal S}(\Phi^A,\Phi_{* A})\}=0 \quad
                                                         \eqno (2.5)
                           $$
  and the initial conditions:
                              $$
 {\cal S}(\Phi^A,\Phi_{* A})=S(\phi)+c^b R_b^A \phi_{*A}+...
                                       \eqno(2.5a)
                            $$
   Where dots means terms containing ghosts and antifields of higher degrees.

  If
                           $$
        [{\bf R_a},{\bf R_b}]=t^c_{ab}{\bf R}_c+E_{ab}^{[AB]}{\cal F}_B
                          $$
where ${\cal F}_A$   are the equations of motion
 (${\cal F}_A={\delta S(\phi) \over \partial \phi^A})$,
 then,
                                $$
 {\cal S}(\Phi^A,\Phi_{* A})=S(\phi)+c^b R_b^A \phi_{*A}+
                   {1\over 2}
              t^c_{ab}c^a c^b c^*_c+
                     {1\over 2}
                     c^a c^b
       E_{ab}^{CD}\phi_{*C}\phi_{*D+}...
                                       \eqno(2.5b)
                            $$

    To the gauge conditions
                          $$
                    f_b=0
                                                        \eqno (2.6)
                           $$
         corresponds the so called "gauge fermion":
                           $$
              \Psi=f_b\nu^b
                                                            \eqno (2.7)
                             $$
  which defines the Lagrangian surface  $\Lambda$ in  ${\cal E}$ by the
           equations
                                 $$
                      F_A(\Phi^A,\Phi_{* A}) =0,
                                                \eqno(2.8)
                                 $$
   where
                                  $$
               F_A=\Phi_{* A}-{\delta \Psi(\Phi) \over \delta\Phi^A}=0
                                                  \eqno (2.9)
                                 $$
   (the  surface embedding in the symplectic space is Lagrangian if it has
 half the di\-men\-sion of  space and the two-form defining the symplectic
structure
is equal to zero on it).
           The partition function $Z$ is given by the integral
    of the master-action exponent over this Lagrangian surface $\Lambda$:
                          $$
   Z=\int e^{{\cal S}(\Phi_A,\Phi^{* A})}
   \delta(\Phi_{* A}-{\delta \Psi(\Phi) \over \delta \Phi^A})
  {\cal D}\Phi^* {\cal D}\Phi
                                                               \eqno (2.10))
                         $$
  (See for details [1]).

  The main statement of the BV formalism is that
 this integral does  not depend on the choise
of the Lagrangian surface $\Lambda$.

 Before going into the geometrical analysis of the formula (2.10)
 we first rewrite it in a more covariant way if the functions $F_a$ which
 define $\Lambda$ by the equation (2.8) are arbitrary.

  It is easy to see that the surface $\Lambda$ defined by (2.8) is Lagrangian
  iff
                                 $$
                        \{F_A,\quad F_B\}\bigg\vert_{F_A=0}=0
                                                        \eqno(2.11)
                                 $$
        Let us consider the integral:
                                     $$
                                     \int
                     e^{{\cal S}(\Phi_A,\Phi^{* A})}
                                 \sqrt{
                                   Ber
                {\delta (G^A,F_B)\over \delta(\Phi^A,\Phi_{* A})}
                                     }
                             \sqrt{
                                   Ber
                               \{G^{\tilde A},F_B\}
                                      }
                                \delta (F)
                       {\cal D}\Phi^* {\cal D}\Phi
                                                                 \eqno (2.12)
                                   $$
         where $G^A$ are arbitrary functions and ${\tilde A}$
  has a parity reversed
   to $A$.

      One can show that if the functions $F_A$ define
    the Lagrangian manifold $\Lambda$ (2.8)  then this integral
 does not depend on the choice of the functions $G_A$ and it does not depend
 on the choice of the functions $F_A$ defining $\Lambda$.
   On the other hand in the case where the functions $F_A$ have the form
  (2.9) and the functions $G^A$ are equal to $\Phi^A$, it evidently coincides
  with the BV integral  (2.10).
                                $$ $$

          \centerline {\bf 3 The survey of BV formalism geometry}

 The formulae (2.5---2.12) of the previous section
have the following geometrical meaning
(see for details [3,5,6] and also [12]).
  In the superspace  $E^{(n.n)}$ with the coordinates

    $z^A=(x^1,...,x^n,\theta^1,...,\theta^n)$
where $x^i$ are even, $\theta^i$ odd coordinates one can consider the
  structure defined by the pair $(\rho , \{\,,\,\})$,
 where $\rho $ is the volume form and
 $\{\,,\,\}$ the odd nondegenerated Poisson bracket
  corresponding to the odd symplectic structure.
   To the structure $(\rho , \{\,,\,\})$
   on
  $ E^{(n.n)}$ corresponds the following geometrical constructions
   which  consistue
 the essence of BV formalism geometry.

 We define
 a second order differential operator on ${ E}$
  (so called $\Delta$--operator)
                          $$
  \Delta_\rho f={1\over 2}div_\rho {\bf D}_f\equiv
  {1\over 2} {{\cal L}_{{\bf D}_f}\rho \over \rho},
                                                  \eqno (3.1)
                         $$
 where ${\bf D}_f$ is the Hamiltonian vector field
 corresponding to the function
$f$. This operator is typical for the odd symplectic geometry[3].
  If $\rho=\rho(z)d^n x d^n \theta$ then
               $$
         \Delta_\rho f=
       {1\over 2\rho}(-1)^{p(A)}{\partial\over \partial z^A}
                 (\rho \{z^A,f\})=
            {1\over 2}\{log \rho,f\}+
       {\partial^2 f\over \partial x^i \partial \theta^i}\,,
                                                       \eqno(3.2)
              $$
  where $p(A)$ is the parity of the coordinate $z^A$.

 We say that the pair  $(\rho , \{\,,\,\})$ is canonical in the
  coordinates

 $z^A=(x^1,...,x^n,\theta^1,...,\theta^n)$
 if $\rho=1\cdot d^n x d^n\theta$  and if the Poisson bracket is
 canonical one:
                 $$
              \{f,g\}=
   {\partial f \over \partial x^i }
   {\partial g \over \partial \theta^i }+
                (-1)^{p(f)}
  {\partial  f\over \partial  \theta^i}
  {\partial  g\over \partial x^i }\,.
                                               \eqno(3.3)
               $$
    Then the $\Delta$--operator takes the canonical expression:
               $$
         \Delta_0 f=
    {\partial^2 f\over \partial x^i \partial \theta^i}\,.
                                               \eqno(3.4)
              $$

  If two $\Delta$--operators
   $\Delta_\rho$ and $\Delta_{\tilde \rho}$
  correspond to two structures with the different
volume forms $\rho$ and ${\tilde\rho}$
 and the same symplectic structure
 \footnote{$^1$}{ Indeed because of Darboux theorem we can always consider
 (at least locally) the canonical symplectic structure (3.3)}.
 then it is easy to see using (3.2) that
              $$
          \Delta_{\tilde \rho}f=
          \Delta_\rho f+ {1\over 2}\{log \lambda,f\}\,,
                                                   \eqno(3.5)
            $$
and
        $$
          \Delta_{\tilde \rho}^2 f=
          \Delta_\rho^2  f+
     \{\lambda^{-{1\over 2}}\Delta_\rho\lambda^{{1\over 2}},f\}\,.
                                                   \eqno3.6)
            $$
  where ${\tilde\rho}=\lambda\rho$.

      For a given structure $(\rho,\{\,,\,\})$
  {\it the following statements are equivalent}:
   \medskip
  i)the operator $\Delta_{\rho} $ is nilpotent
                          $$
                    \Delta_{\rho}^2=0\,,
                                                 \eqno(3.7i)
                          $$
 ii)
 the function $\rho(z)$ defining the volume form $\rho$  obeys
the equation:
                 $$
             \Delta_0  \sqrt \rho=0
                                  \eqno   (3.7ii)
              $$
 iii)
      there exist  coordinates in which the pair
  $(\rho , \{\,,\,\})$ is canonical.
   \footnote{$^2$}{ The structures $(\rho , \{\,,\,\})$ for which
these properties are obeyed are called SP structures [5].
 One of us (O.M.K.) wants to note that in  [3] where was first
 introduced the $\Delta$--operator related to the structure
 $(\rho , \{\,,\,\})$ for an arbitrary volume form in
 superspace the false statement that every
$(\rho , \{\,,\,\})$ structure is SP structure was made}

   The iii)$\Rightarrow$i) is evident, the i)$\Leftrightarrow$ii)
immediately follows from (3.6). The i)$\Rightarrow$iii)
needs more detailed analysis.
        \medskip

  The pair $(\rho , \{\,,\,\})$  generates  the invariant volume form
 $\rho_\Lambda$ on arbitrary  Lagrangian manifolds $\Lambda$
  in ${ E}$---"the square root of the volume
form $\rho$" in the following way [5]:
                       $$
   \rho_\Lambda(e_1,\cdots,e_n)=
    \sqrt{\rho(e_1,\cdots,e_n,f_1\cdots,f_n)}
                                        \eqno (3.8)
                        $$
where $\{e_i\}$ are the vectors tangent to  $\Lambda$
 and $\{f_i\}$ are arbitrary vectors such that
                        $$
              w(e_i,f_j)=\delta_{ij}\,.
                         $$

  In these   terms the BV formalism has the following geometrical meaning:
 We consider in the superspace ${\cal E}$ of the fields and antifields the pair
 $(\rho , \{\,,\,\})$ where the volume form is defined by the master-action:
                              $$
    \rho=e^{2 {\cal S}},
                                   \eqno(3.9)
                            $$
     and  $ \{\,,\,\}$ is defined by(2.3). Then using
   i), ii), iii) and comparing  formulae (3.7) with formulae
 (2.3-- 2.5) we see  that the master--equation is nothing but
 the condition of nilpotency of the corresponding $\Delta$ operator.
 The partition function is nothing but the integral of the invariant volume
   form (3.8) on the
  Lagrangian surface $\Lambda$ [5] and the eq. (2.12)
 is the covariant expression for this
 volume form.

  In the next section we will try to understand these statements from the point
of view
of integration theory on surfaces.

  \vfill\eject
                   \medskip

    \centerline  {\bf 4 Integration over surfaces}
                           \medskip
In this section we present the basic objects of  integration theory
on  supermanifolds: densities and dual densities [8--10].
 We consider the special class of densities on which
the exterior derifferential can be defined
 correctly---pseudo\-dif\-ferential forms [7--10].
 Then we describe the Baranov--Schwarz (BS) representation of the
pseudodifferential
forms via the function on the superspace associated to the tangent
bundle of initial space [8]. Considering the dual construction we
  show that the closure of the pseudodifferential
 form in the BS representation
is formulated in  terms of the $\Delta$ operator.
                    \medskip
     \centerline  {\it Densities}

     Let $\Omega$ be an arbitrary
supersurface in the superspace $E$ with coordinates $z^a$,
      given by
   a  parametrization $z^a=z^a(\zeta^s)$.
     The function $L(z^a,{\partial
  z^a\over \partial \zeta^s})$ on $E$ is  called a density (covariant density),
 if
 is satisfies  the condition [9]:
                          $$
      L(z^a,{\partial z^a\over \partial
\zeta^{s^\prime}} K^{s^\prime}_s)= L(z^a,{\partial z^a\over \partial \zeta^s})
	{\rm Ber}K^{s^\prime}_s ,
                                                  \eqno (4.1)
                         $$
     where $Ber$ is
  the superdeterminant of the matrix.

     Then the following integral does not
 depend on the choice of the parametrization of the surface $\Omega$
                         $$
    \Phi_\Omega (L)=
\int  L(z^a(\zeta),{\partial z^a(\zeta)\over \partial
   \zeta^s})d\zeta   ,                              \eqno(4.2)
                         $$
 and  correctly defines the functional on the surface $\Omega$
 corresponding to the density $L$.

 In the bosonic case where there are not odd variables, one can see
that if a density $L$ is a linear function of the
 ${\partial z^a(\zeta)\over \partial
   \zeta^s}$  then to $L$ corresponds a differential form.
 The covariant density is closed if it  satisfies identically
   the condition:
                             $$
         \Phi_{\Omega+\delta \Omega}({L}) =
               \Phi_{\Omega}({L})
                                                     \eqno(4.3)
                             $$
 for an arbitrary variation of an arbitrary
  surface $\Omega$ (up to boundary terms).

 It is easy to see that
                            $$
         \Phi_{\Omega+\delta \Omega}(L) -
               \Phi_{\Omega}(L)=
                       {\cal F}_a(z)\delta z^a
                                                      \eqno (4.4)
                            $$
where
                            $$
                 {\cal F}_a(z)={\partial L\over \partial z^a}-
                      (-1)^{p(a)p(s)}
                {d \over d\zeta^s}{\partial L\over \partial z^a_{,s}}
                                                          \eqno(4.5)
                            $$
 are the left part of the Euler-Lagrange equations of
 the functional $\Phi$(L).
   \medskip
{\it   How to define exterior derivative  operator on the densities}?

 If  $d$ is the exterior derivative , then
                      $$
         \Phi_{\Omega+\delta \Omega}({L}) -
               \Phi_{\Omega}({L})=
               \Phi_{\delta V}({d L})
\quad {\rm (Stokes}\quad {\rm theorem)}.
                                                   \eqno (4.6)
                       $$

     Eq.(4.6) put strong restrictions on the class of densities
 on which the operator $d$ is correctly defined [10].
    Comparing (4.4), (4.5) and (4.6) we see that $d$ is correctly defined
    if ${\cal F}_a(z)$ in (4.5)
     {\it do not contain the second derivatives of
     $\zeta$} [10]:
                  $$
       {\partial^2 L\over {\partial z^a_{,s} \partial
	   z^b_{,t}}}=
  -(-1)^{p(s)p(t) + (p(s)+p(t))p(b)}
    {\partial^2
	   L\over {\partial z^a_{,t} \partial z^b_{,s}}}.
                                      \eqno (4.7)
                  $$
   In this case $dL$ defined by (4.6)
  does not depend on the second derivatives and
                                    $$
                           d^2=0\quad.
                                         \eqno (4.8)
                                   $$

The densities, which obey  the conditions (4.7)
  are called pseudodifferential forms.

  In the bosonic case from (4.7) follows that the density is a linear
 function of the variables
   ${\partial z^a(\zeta)\over \partial
   \zeta^s}$ i.e. the exterior derivation can be defined only on the densities
 which correspond to the differential forms. In the supercase in general
 from (4.7) linearity conditions do not follow---the differential forms
 in the superspace are not in general  integration objects over supersurfaces.
 It is the pseudodifferential forms which take their place
  as integration objects
  obeying  Stokes theorem [7--10]).

To
obtain  the pseudodifferential forms, Baranov and Schwarz in [8]
suggested the following procedure
  which seems very natural in the spirit of a ghost technique:

Let $STE$ be the superspace
      associated to the tangent bundle $TE$ of the superspace $E$
   and    $(z^a, z^{ * a})$   its (local) coordinates.
 The coordinates $z^{*a}$ transform from map to map like $dz^a$,
  and their parity is reversed: $ p(z^{* a})=p(z^a)+1$.
  Then to
  an arbitrary function $W(z^a, z^{* a})$ on $STE$
   corresponds the density:
                     $$
                      L_W=
  L(z^a,{\partial z^a\over \partial \zeta^s})=
  \int W(z^a, {\partial z^a\over \partial \zeta^s}\nu^s)d\nu
   \quad,
                                  \eqno (4.9)
                         $$
 where $\nu^s$
      has the reversed parity:
                     $$
    p(\nu^s)=p(\zeta^s )+1  .
                                   \eqno (4.10)
                      $$
 It is easy to see using (4.10) that (4.9) obeys
 equations (4.1) and (4.7)
  so that equation (4.9) indeed defines a density which is a
 pseudodifferential form.
   We say that the function $W$ is the BS representation of the
 pseudodifferential form $L_W$.

   A simple calculations show  that in the BS representation
 the exterior differentiation
operator has the following expression:
                  $$
            {\hat d}=(-1)^{p(a)}z^{*a}{\partial\over \partial z^a}
                                   \eqno (4.11)
                  $$
                    $$
              (d(L_W)=L_{{\hat d} W})\,.
                    $$

              \vfill\eject
   \centerline {\it Dual densities}
            \medskip
   Consider now the dual constructions.

 Let $E$ be the superspace , and
   $\rho(z)dz$  the volume form is  defined on it.

         Let $\Omega$ be an arbitrary
supersurface in the superspace $E$ with coordinates $z^a$,
      given not by
    the parametrization $z^a=z^a(\zeta^s)$ but by the equations
                     $$
                f^\alpha(z)=0\,.
                                       \eqno (4.12)
                      $$
    The function  ${\tilde L}= {\tilde
 L}(z^a,{\partial f^\alpha \over \partial z^a})$ is called  a D-density
    (dual
    density) if it is satisfied to the condition   :
                     $$
        {\tilde L}(z^a,{\partial f^\alpha (z)\over \partial
  z^a} \eta^\beta_{\alpha})= {\tilde L}(z^a,{\partial f^\alpha(z)\over
	 \partial z^a}) {\rm Ber}\;\eta^\beta_{\alpha}\,.
                                           \eqno (4.13)
                      $$

     Then the following integral does not
 depend on the choise of the equations (4.12) which define
 the surface $\Omega$
                         $$
    \Phi_\Omega (\tilde L)=
\int  {\tilde L}(z^a\,,{\partial f^\alpha(z)\over \partial z^a})
                        \delta (f^\alpha (z))\rho(z)dz\,,
                                                \eqno(4.14)
                         $$
 and  correctly defines the functional on the surface $\Omega$
 corresponding to the D--density ${\tilde L}$.

The D-density ${\tilde L}$ corresponds to the density $L$
  (${\tilde L}\rightarrow L$) if for the arbitrary surface
 $\Omega$ the functionals
  (4.2) and (4.14)
  coincide. (See for the details [9])

   (For example the integrand in (2.12) is a D--density which
correspond to the density (3.8))

  The D-density is closed, if
it satisfies the  condition (4.3) (where we replace
  ${\tilde L}\rightarrow{ L}$).

  One
can obtain the dual densities corresponding to pseudodifferential forms
  (such densities are called pseudointegral forms) by the
procedure dual  to the Baranov-Schwarz one:

Let $ST^*E$ be the superspace
      associated to the cotangent bundle $T^*E$ of the superspace $E$
   and    $(z^a, z^*_ a)$  its (local) coordinates.
 The coordinates $z^*_a$ transforms from
 map to map like ${\partial\over \partial z^a}$,
  and their parity is reversed: $ p(z^* _a)=p(z^a)+1$.
  Then  to
  an arbitrary function $W(z^a, z^*_a)$ on $ST^*E$
   corresponds the D--density---pseudointegral form:
                         $$
                           {\tilde L}_W=
 {\tilde L}(z^a,{\partial f^\alpha\over \partial z^a})=
\int W(z^a,{\partial f^\alpha\over \partial z^a}\nu_\alpha )d \nu\,.
                                                        \eqno(4.15)
                         $$
   where $\nu^\alpha$
      have the reversed parity like in (4.10):
                     $$
    p(\nu^\alpha)=p(f^\alpha)+1  .
                      $$

  The functional (4.14) can be  expressed in term of the function $W$
 in the following way:
                        $$
           \Phi_\Omega (\tilde L)=
                       \int
                \rho(z)W(z,z^*)
   \delta(z^*_a-{\partial f^\alpha\over\partial z^a}\nu_\alpha)
                 \delta (f^a)
                     dzdz^*d\nu\,.
                                                  \eqno(4.15a)
                       $$
 A straightforward calculation show  that the operator
 of exterior differentiation
 ${\hat {\tilde d}}$
  in the BS representation of the pseudointegral forms
   has the following expression:
                         $$
                   {\hat {\tilde d}}=
                  {1\over \rho}
           {\partial \rho \over \partial z^a}
           {\partial \over \partial z^*_a}+
           {\partial^2 \over \partial z^a \partial z^*_a}\,.
                                                         \eqno(4.16)
                       $$
                 \centerline{
    (If ${\tilde L}={\tilde L}_W \rightarrow L$ then
        ${\tilde L^\prime}=
        {\tilde L}_{{\hat {\tilde d}}W} \rightarrow dL$).}

                      \medskip
 Comparing the equations (4.16) and (3.2), we see that
  on the superspace $ST^*E$  it is natural to consider
 the structure $({\hat \rho},\{\,,\,\})$
 (See the Sect.3)
where  $\{\,,\,\}$ is the canonical odd symplectic
structure on $ST^*E$ generated by the relations
                   $$
         \{z^a,z_b\}=\{z^{*a},z^*_b\}=0,\quad
          \{z^a,z^*_b\}=(-1)^{p(a)}\delta^a_b
                                                       \eqno(4.17)
                   $$
      and the volume form
                   $$
          {\hat \rho}
       =\rho^2(z^1\cdots z^n)dz^1...dz^n dz^*_1...dz^*_n\,.
                                                   \eqno(4.18)
                  $$
   (One can note that (4.18) is in the accordance with
 (3.8).---The space $E$ with volume form $\rho$ is
 evidently the Lagrangian surface in $ST^*E$ with
 volume form (4.18))

   Comparing  (4.16) and (3.2) we see that
 to the operator of the exterior differentiation
 corresponds the $\Delta$--operator:
                  $$
            {\hat {\tilde d}}= \Delta_{\hat \rho}
                                                   \eqno (4.19)
                  $$
   and the condition of closure
 of the dual density ${\tilde L}_W$ in the BS representation is
                             $$
                 \Delta_{\hat \rho}W=0\,,
                                               \eqno(4.19a)
                            $$
  where  ${\hat \rho}$ is defined by
   (4.18) and $\Delta_{\hat \rho}$ by (3.2). This operator in this case is
nilpotent
because it corresponds to exterior differentiation operator.
 (Independently from (4.16) and (4.8) it follows
from (4.18) and (3.7ii) or from
(4.18) and (3.7iii) because ${\hat \rho}$ depends on the
half of the variables of the superspace $ST^*E$.)

                            $$ $$
                            $$ $$

    \centerline {\bf 5 The closed densities and the BV formalism geometry.}
                           \medskip

   In this section we consider two examples of the previous
constructions comparing them with the constructions of the
  Sections 2,3 and 4.
 We check connections between the gauge symmetries of the theory,
 the densities which are integrand in the partition function after
eliminating gauge degrees of freedom, and
 volume forms obeying the BV--master--equation.

   \medskip
 {\bf Example 1}
 Let $R^a(z^a){\partial \over \partial z^a}$ be an even vector field
on the superspace $E$ with coordinates $(z^1,\dots,z^n)$
 and with volume form $\rho=\rho(z)dz^1\cdot\cdot\cdot dz^n$.
 To this vector field corresponds the D--density
                        $$
      {\tilde L}=R^a(z^a){\partial f\over \partial z^a}\,.
                                         \eqno (5.1)
                        $$

  One can define the functional on the surfaces of codimension
$(1.0)$  corresponding to the density (5.1):
                        $$
       \Phi_\Omega (\tilde L)=
     \int{\tilde L}(z^a\,,{\partial f(z)\over \partial z^a})
                        \delta (f(z))\rho(z)dz\,=
     \int R^a(z){\partial f(z)\over \partial z^a}
                       \delta (f (z))\rho(z)dz\,,
                                                    \eqno (5.2)
                      $$
 where  $f=0$ is the equation which defines the surface
  $\Omega$ ($f$ is an even function).
This functional is nothing but the well--known formula for the
 flux of the vector field through the surface $\Omega$.
 It is evident that the density ${\tilde L}$ in (5.1)
is pseudointegral form. To this density corresponds
  the function (4.15)
                   $$
            W=(-1)^{p(a)}R^a(z)z^*_a
                                                \eqno(5.3)
                   $$
   on $ST^*E$. (${\tilde L}={\tilde L}_W$).
The condition of closure of the density (5.1) is the Gauss formula:
                  $$
                  div_\rho {\bf R}=
              {1\over \rho}(-1)^a
             {\partial (\rho R^a)\over \partial z^a}=
                       0\,.
                                                \eqno(5.4)
                   $$
   In BS representation it is(4.18, 4.19)
                   $$
  \Delta_{\hat \rho}W=
        \Delta_{\rho^2} W=0.
                                              \eqno(5.5)
                 $$
  We can consider this example as a toy example of field theory.

  Let a space $E$ be the space of fields configurations
    $(z^a\rightarrow \varphi^a)$
Let $R^a(z){\partial \over \partial z^a}$ be the
 "gauge" symmetry
of the action $S(z)$ (compare with (2.1)):
                 $$
           R^a(z)
   {\partial S(z)\over \partial z^a}=0
                                                \eqno(5.6)
                 $$
 and this symmetry preserves the canonical volume form:
                $$
             (-1)^{p(a)}
  {\partial  R^a\over \partial z^a}=
                       0\,.
                                                \eqno(5.7)
                $$
     If we put
              $$
           \rho=e^S
                                               \eqno(5.8)
              $$
   then we see that the functional (5.2) corresponding to the
density (5.1) constructed via the "gauge symmetry"
 ${\bf R}$ is the partition function of the theory with the action $S$
after eliminating the "gauge" degrees of freedom corresponding
to the symmetry ${\bf R}$.  From (5.6--5.8)) follow (5.4, 5.5) hence
(5.1) is closed and (5.2) is "gauge" independent.
                       $$ $$
 Now we consider the more realistic
          \medskip
  {\bf Example 2} Let
                          $$
    \{{\bf R}_\alpha=R^a_\alpha(z){\partial\over\partial z^a}\},\,\,
                 (\alpha=1,\cdots,m)
                                                  \eqno(5.9)
                           $$
be the collection of the vector fields on the superspace
 $E$ with coordinates

 $(z^1,\dots,z^n)$
 and with volume form
                $$
          \rho=\rho(z)dz^1\cdot\cdot\cdot dz^n\,.
                                                  \eqno(5.10)
                  $$
 To (5.9) corresponds D--density
                      $$
           {\tilde L}=
      Ber(R^a_\alpha(z){\partial f^\beta\over\partial z^a})\,
                                                       \eqno(5.11)
                      $$
 (the condition (4.13)  is evidently satisfied.)
   One can consider the functional:
                      $$
                 \Phi_\Omega({\tilde L})=
                   \int
    Ber(R^a_\alpha(z){\partial f^\beta\over\partial z^a})\,
                     \rho (z)
                  \delta(f) d^nz
                                                      \eqno(5.12)
                       $$
  where $\Omega$ is the surface defined by the equations
                      $$
                   f^\alpha=0\,.
                      $$
 (In the usual (not super)case,
 (5.12) can be considered as the flux of the polivectorial
 field

 ${\bf R_1}\wedge\cdots\wedge{\bf R_m}$ through the surface $\Omega$.)

  One can see that ${\tilde L}$ in (5.11) is pseudointegral form
 ${\tilde L}_W$ where the $W$---BS representation of this density
 can be defined by the following formal relation:
                        $$
                        W
                        =
                        \int
               e^{c^\alpha R_\alpha^a(z)z^*_a}dc
                                                             \eqno(5.13)
                       $$
 where we introduce additional variables
  (ghosts) $c^\alpha$ ($p(c^\alpha)=p(\nu^\alpha)$).
  ((5.13) is correct if all the symmetries
 ${\bf R}_\alpha$ are even).

 Let the equations (5.6), (5.7) be satisfied for all ${\bf R}_\alpha$---
  these vector fields being the gauge symmetries of the theory with the
 action $S$. Again as in the Example 1 we consider as volume form the
 exponent of the action (5.8). Does the density (5.11) is closed in this case?

   It is easy to see that
                       $$
                   \forall \,\,
                     R^a(z)
       {\partial S(z)\over \partial z^a}=0
                 {\bf     \rightarrow}
             [{\bf R_\alpha},{\bf R_\beta}]=
              t^\gamma_{\alpha\beta}{\bf R}_c+
               E_{\alpha\beta}^{[ab]}
      {\partial S(z)\over \partial z^b}\,.
                                                            \eqno(5.14)
                    $$
  To check the relation with the BV--formalism
 we consider instead superspace $E$ the superspace $E^e$ enlarged
 with the additional coordinates $c^\alpha$.
 (The coordinates of $E^e$ are
 $z^A=(z^a, c^\alpha)$).
  The volume forms $\rho(z)$
on  $E$ and ${\hat \rho}$ on  $ST^*E$ (see (4.18))
 and the symplectic structure (4.17) are naturally
 prolongated on  $E^e$ and $ST^*E^e$.

   Using (4.15, 4.15a) and (5.13) we rewrite (5.12) as the integral over
the space $T^*SE^e$:
                   $$
        \Phi_\Omega({\tilde L})=
                  \int
                { \rho}
        e^{c^\alpha R_\alpha^a(z)z^*_a}dc
   \delta(z^*_a-{\partial f^\alpha\over\partial z^a}\nu_\alpha)
                 \delta (f)
                  dzdz^*d\nu\,=
                                                        \eqno (5.15)
                  $$
                  $$
                   \int
                { \rho}
              W^e(z^A,z^*_A)
   \delta(z^*_A-{\partial f^\alpha\over\partial z^A}\nu_\alpha)
                     dzdz^*d\nu\,,
                                                    \eqno(5.16)
                 $$
 where
                    $$
      W^e(z^A,z^*_A)=e^{c^\alpha R_\alpha^a(z)z^*_a}
                                                     \eqno(5.17)
                   $$
   is the BS representation of
the pseudointegral form in $ST^*E^e$. Using  (4.19) we can check
 its closure.

   ((5.15, 5.16)   is the partition function of the theory obtained after
performing the Fa\-de\-ev--Popov trick).

  Let
                     $$
          \Delta_{\hat \rho}W^e=
                   \Delta_{\hat \rho}
             e^{c^\alpha R_\alpha^a(z)z^*_a}
                   = 0\,
                                                       \eqno (5.18)
                    $$
 be satisfied.  The condition
 (5.18) means that  not only the function
$W$ on $ST^*E$ corresponds to the closed density on $E$
 ( i.e.the partition function (5.15) is gauge invariant) but
 the function
 $W^e$ on $ST^*E^e$ corresponds to the closed density on $E^e$ as well.
 In this case from (3.6) and (3.7) follows
that the $\Delta$ operator corresponding to the volume form
                    $$
        {\hat \rho}\prime=  {\hat \rho}\cdot(W^e)^2
                                                      \eqno (5.19)
                   $$
   is nilpotent ($W^e$ in contrary to $W$ is even)
as well as the $\Delta$ operator corresponding to
  the volume form (5.8).
   Now  from (3.7) follows that   the master-action ${\cal S}$

 related with ${ \rho}\prime$ in the same way as $S$ is related
with ${ \rho}$ in (5.8):
                    $$
           {\cal S}=S+c^\alpha R_\alpha^a z^*_a\,,\qquad
          ({ \rho}\prime=e^{\cal S})
                                                      \eqno (5.20)
                     $$
    obeys the master-equation.
 So in the case where (5.18) holds,
  starting from gauge symmetries we constructed
the closed density (5.12, 5.13), interpreting the volume
 form as the exponent of the action.
The corresponding functional (5.12) is
 the partition function.  Localizing this density in the
 space enlarged with the ghosts we came to the volume form
(exponent of the master action) which obeys to the master-action.

     In general case the density (5.11) {\it is not closed} and
    the partition function (5.12, 5.16) is not gauge invariant.

     Even in the case where the algebra of the symmetries is closed:
                      $$
           t^\gamma_{\alpha\beta}=const\,\,\,
                    {\rm and}\,\,\,
                     E_{\alpha\beta}^{[ab]}\equiv 0
                                                         \eqno(5.21)
                        $$

   the application of the $\Delta$--operator  (4.19) to
  (5.17) and(5.13) give us
                       $$
                \Delta_{\rho^2}W^e=
              \Delta_{\rho^2}
          e^{c^\alpha R_\alpha^a(z)z^*_a}dc=
                    {1\over 2}
                   c^\alpha c^\beta
               (t^\gamma_{\alpha\beta}
                      R^a_\gamma(z)+
                    E_{\alpha\beta}^{[ab]}
      {\partial S(z)\over \partial z^b})z^*_a
         e^{c^\alpha R_\alpha^a(z)z^*_a}dc\,
                                                        \eqno(5.22)
                      $$
and
                     $$
                \Delta_{\rho^2}W=
              \Delta_{\rho^2}
                     \int
          e^{c^\alpha R_\alpha^a(z)z^*_a}dc=
                      \int
                     c^\alpha c^\beta
               (t^\gamma_{\alpha\beta}
                      R^a_\gamma(z)+
                    E_{\alpha\beta}^{[ab]}
      {\partial S(z)\over \partial z^b})z^*_a
         e^{c^\alpha R_\alpha^a(z)z^*_a}dc\,.
                                                        \eqno(5.23)
                      $$

 In particular it is easy to see
from (5.22) that if
 the algebra of the symmetries is abelian we come
to (5.18).

 If, for example,
  the symmetries are even
 and they consist the closed unimodular algebra
  (  $E_{\alpha\beta}^{[ab]}=0$,
  $t_{\alpha\beta}^\gamma=const$ and
        $\sum_\alpha t^\alpha_{\alpha,\beta}=0$)
              then the right hand side of (5.23)
 is vanishing,so the function $W$ corresponds
 to closed density in $E$ (i.e.the partition function
is gauge invariant). But  the function $W^e$ in (5.22)
 does not correspond
to closed density in $E^e$. To  close it in this case
 one have to consider in the space $ST^*E^e$ the function
                        $$
          W^{e\prime}(z^A,z^*_A)=
          e^{c^\alpha R_\alpha^a(z)z^*_a
              +{1\over 2}t_{\alpha\beta}^\gamma
          c^\alpha c^\beta c^*_\gamma}
                        $$
which corresponds to a closed density in $E^e$. So the corresponding
   volume form  and the master action
                              $$
          {\cal S}=S+c^\alpha R_\alpha^a z^*_a
       +{1\over 2}t_{\alpha\beta}^\gamma
          c^\alpha c^\beta c^*_\gamma
                     $$

                   obey the master equation  (compare with (2.5b).

 In the general case the density (5.11, 5.13)  plays the role of initial
conditions for constructing the closed density in enlarged space--i.e.
the volume form (the exponent of the master--action)
   obeying the (3.7).
      $$ $$ $$ $$
    \centerline {\bf Acknowledgments}
   $$ $$
 We are grateful to I.A.Batalin and to I.V.Tyutin
 for fruitful discussions.

 One of us (O.M.K.) is grateful to O.Piguet
  and other colleagues from the
 Department of Theoretical Physics of Geneva University.
Their hospitality made it possible to finish this work.
 The work of A.N. has been made possible by
a fellowship of Grant No. M21300 and INTAS Grant
93--2494 and is carried out within the research
programm of International Center for
Fundamental Physics in Moscow.

           \vfill\eject

         \centerline {\bf References}

1] Batalin I.A., VIlkovisky G.A.
  --Phys. Lett., {\bf 102B}, 27 (1981);

    Phys.Rev {\bf D28} 2567 (1983);
    Nucl.Phys. {\bf B234}  106 (1984 )
      \medskip
[2] E. Witten---Mod. Phys. Lett. {\bf A5} 487 (1990)
                 \medskip

[3] Khudaverdian O.M---J. Math. Phys., {\bf 32} 1934 (1991):

 Preprint of Geneva Universite UGVA-DPT 1989/05--613 .
                 \medskip

 [4] M. Henneaux
--Phys. Lett., {\bf B315}, 283 (1993)

 H.Hata, B. Zwiebach---Annals of Phys., {\bf 229}, 177 (1994)

  J.Alfaro, P.H.Damgaard---Nucl.Phys., {\bf B404}, 751 (1993)
                 \medskip

 [5] Schwarz A.S. - Comm. Math. Phys. {\bf 155},249 (1993)
                 \medskip

 [6] O. M. Khudaverdian , A. P.  Nersessian--- Mod. Phys. Lett
{\bf A8}, 2377 (1993);

    J.Math.Phys., {\bf 34} 5533 (1993)
                 \medskip

 [7]  J.N. Bernstein, D.A.Leitec --- Funk.Anal.i ego pril.
   {\bf 11},No1, 55 (1977);

   {\it ibid} {\bf 11}, No3, 70 (1977).
                 \medskip

 [8]  M. Baranov, A. S. Schwarz -- Funk. anal i ego  pril.
  {\bf 18}, No.2 ,53  (1984);

   {\it ibid} {\bf 18}, No.3, 69 (1984)
                 \medskip

 [9] A.Gayduk, O.M. Khudaverdian, A.S.Schwarz--- Teor. i Mat. Fizika
     {\bf 52}, 375 (1982)
                 \medskip

 [10]  T. Voronov  - Geometric integration theory
over supermanifolds.---

 Sov. Scient. Rev.C Math.Phys., Vol.9, pp.1--138 (1992)
                 \medskip

 [11] I.A. Batalin, I.V. Tyutin---
  Mod.Phys.Lett., {\bf A8}, 283 (1993); {\it ibid}
     {\bf A8}, 3673 (1993); {\it ibid}
  {\bf A9}, 1707 (1994)
   (Preprint FIAN/TD/4--94)
        \medskip
 [12] I.A. Batalin, I.V. Tyutin---Int. J. Mod.Phys. {\bf A8}
  2333 (1993)

  \bye